\documentstyle[aps]{revtex}
\input epsf
\def\jepsfbox#1{\typeout{#1} \epsfbox{#1}}
\def\plotonesc#1#2{\begin{center} \leavevmode
\epsfxsize=#2\columnwidth \jepsfbox{#1} \end{center}}
\def\jcite#1#2{#1 \cite{#2}}
\def\tilde{\mathaccent"365}			
\def\etal{{\it et al.\ }}
\def\eg{{\it e.g.~}}
\def\ie{{\it i.e.~}}

\def\rmmat#1{{\hbox{\rm #1}}}
\def\rmscr#1{\rmmat{\scriptsize #1}}
\newcommand{\comment}[1]{{\relax}}
\newcommand{\be}{\begin{equation}}
\newcommand{\ee}{\end{equation}}
\newcommand{\ba}{\begin{eqnarray}}
\newcommand{\ea}{\end{eqnarray}}
\def\p{\partial}
\def\d{{\rm d}}

\def\dd#1#2{\frac{\d #1}{\d #2}}
\def\pp#1#2{\frac{\p #1}{\p #2}}
\def\figref#1{Fig.~\ref{fig:#1}}
\def\eqref#1{Eq.~\ref{eq:#1}}
\begin{document}
\draft
\newcommand{\bfi}{{\bf B}} \newcommand{\efi}{{\bf E}}
\newcommand{\lag}{{\cal L}} \newcommand{\dLIII}{{\frac{\partial^3
\lag}{\partial I^3}}} \newcommand{\dLII}{{\frac{\partial^2
\lag}{\partial I^2}}} \newcommand{\dLI}{{\frac{\partial \lag}{\partial
I}}} \newcommand{\dLKKK}{{\frac{\partial^3 \lag}{\partial K^3}}}
\newcommand{\dLKK}{{\frac{\partial^2 \lag}{\partial K^2}}}
\newcommand{\dLK}{{\frac{\partial \lag}{\partial K}}}
\newcommand{\dLIK}{{\frac{\partial^2 \lag}{\partial I \partial K}}}
\title{Electromagnetic Shocks in Strong Magnetic Fields}
\author{Jeremy S. Heyl
\thanks{Current address: Theoretical Astrophysics, mail code 130-33, 
California Institute of Technology, Pasadena, CA 91125}
\and Lars Hernquist\thanks{Presidential Faculty
Fellow}}
\address{Lick Observatory,
University of California, Santa Cruz, California 95064, USA}
\maketitle
\begin{abstract}
We examine the propagation of electromagnetic radiation through a strong
magnetic field using the method of characteristics.  Owing to
nonlinear effects associated with vacuum polarization, such waves can
develop discontinuities analogous to hydrodynamical shocks.  We derive
shock jump conditions and discuss the physical nature of these
non-linear waves
\end{abstract}

\section{Introduction}

The nonlinear properties of electromagnetic waves traveling through a
magnetized vacuum is of particular interest in the study of neutron
stars.  \jcite{Heisenberg and Euler}{Heis36} and
\jcite{Weisskopf}{Weis36} first derived nonlinear corrections to the
Maxwell equations of the electromagnetic field.  \jcite{Lutzky and
Toll}{Lutz59} and \jcite{Zheleznyakov and Fabrikant}{Zhel82} applied
the weak-field expansion to show that shock waves can develop in the
electromagnetic field.  \jcite{Bialynicka-Birula}{Bial81} used the
full expression for the nonlinear correction to the Lagrangian to
study the generation of harmonics and other nonlinear phenomena in the
propagation of EM radiation.

In this paper, we use the method of characteristics to study the
evolution of waves governed by an arbitrary Lagrangian and then apply
these techniques to the Heisenberg-Euler-Weisskopf-Schwinger
Lagrangian \cite{Heis36,Weis36,Schw51}. We find a concise expression
for the opacity to shocking for disturbances travelling through an
arbitrarily strong mangetic field.  After deriving a form
for the energy-momentum tensor for an arbitrary
electromagnetic Lagrangian, we use relativitistic fluid mechanics to
derive the shock jump conditions and the long-term evolution of
electromagnetic disturbances propagating in magnetic fields.

\section{Deriving the Characteristics}

\begin{figure}
\plotonesc{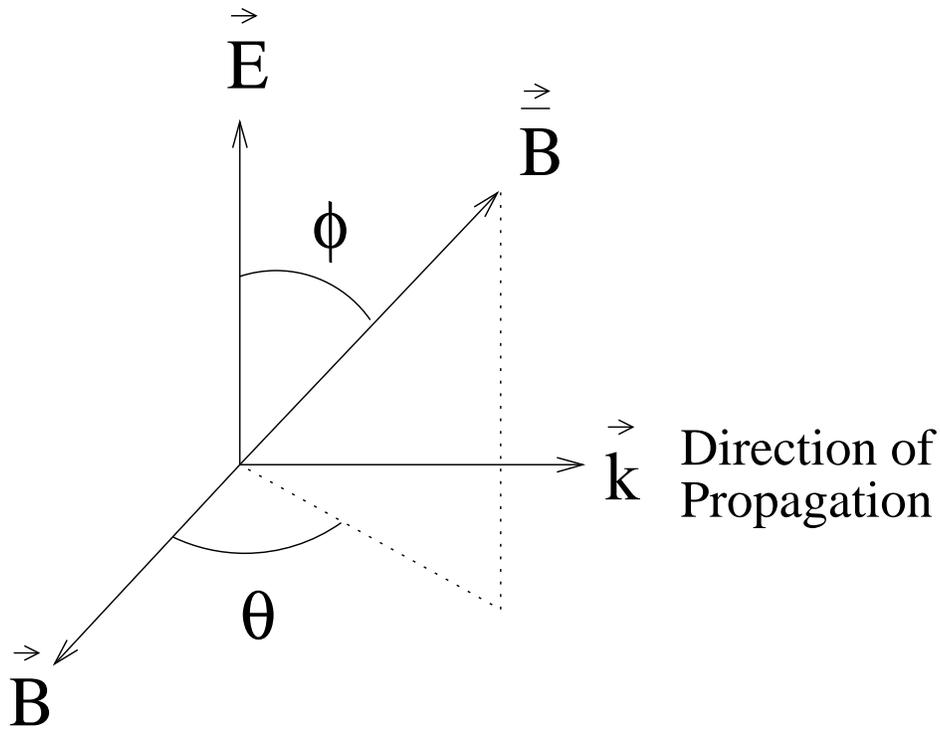}{0.7}
\caption{Illustration of the field geometry}
\label{fig:geometry}
\end{figure}

We will use the method of characteristics to study the evolution of a
disturbance of the electromagnetic field.  In general,
the relativistic Lagrangian ($\lag$) of the electromagnetic 
field is a function of the two invariants of the field.  We follow the
notation of \jcite{Lutzky and Toll}{Lutz59} and 
\jcite{Heisenberg and Euler}{Heis36} and define
\be
I = F_{\mu\nu} F^{\mu\nu} = 2 \left ( |\bfi|^2 - |\efi|^2 \right )
\ee
and
\be
K = - \left (\frac{1}{2} \epsilon^{\lambda\rho\mu\nu}
F_{\lambda\rho} F_{\mu\nu} \right)^2 =
	- (4 \efi \cdot \bfi )^2.
\ee
As illustrated in \figref{geometry}, we choose coordinates so that the
radiation is polarized in the $z$-direction and travels along the
$y$-axis toward positive $y$.  The ambient magnetic field makes an angle
$\phi$ with the electric field, and the projection of the magnetic field
into $x-y$ plane makes an angle $\theta$ with respect to the $x$-axis
(magnetic field of the wave).

With these definitions, the invariants are
\begin{eqnarray}
I & = & 2 \left ( ({\bar B} \cos \theta \sin \phi + B )^2 + ({\bar B} \sin \theta
\sin \phi)^2 + ({\bar B} \cos \phi)^2- E^2 \right ) \\
K & = & -(4 E {\bar B} \cos \phi)^2
\end{eqnarray}
where ${\bar B}$ is the strength of the ambient magnetic field, $E$ and $B$ 
are the strengths of the electric and magnetic fields associated with the 
radiation.

\comment{
\subsection{$\epsilon, \eta$ formalism}

We introduce the new coordinates
\ba
\epsilon &=& y + t \\
\eta &=& y - t
\ea
where we set $c\equiv 1$.  We characterize the traveling wave by a vector 
potential with one non-zero component,
\be
A_z = \psi(y,t) = \psi(\epsilon,\eta).
\ee
We have assumed that the vector potential has only one independent
component which allows us to treat the problem using characteristics.
Unfortunately, we cannot follow the interaction between two
polarizations which was treated by \jcite{Bialynicka-Birula}{Bial81}.

Using the definition of the vector potential we get
\ba
E &=& \psi_\eta - \psi_\epsilon \\
B &=& \psi_\eta + \psi_\epsilon
\ea
where we have used subscripts to denote partial differentiation.  
Substituting this into the definitions of $I$ and $K$, we get
\begin{eqnarray}
I & = & 2 (4 \psi_\eta \psi_\epsilon + {\bar B}^2 + 2 (\psi_\eta + \psi_\epsilon) {\bar B}
\cos \theta \sin \phi) \\
K & = & -16 (\psi_\eta - \psi_\epsilon)^2 {\bar B}^2 \cos^2 \phi.
\end{eqnarray}

In the $(\eta,\epsilon)$ coordinates Hamilton's principle assumes the
simple form \cite{Lutz59}
\be
{\partial \over \partial \epsilon} \left ( \pp{\lag}{\psi_\epsilon} \right ) +
{\partial \over \partial \eta} \left ( \pp{\lag}{\psi_\eta} \right ) = 0.
\ee
Since the Lagrangian is a function of $I$ and $K$ alone, we can
rewrite this in terms of derivatives of $\lag$ with respect to $I$ and
$K$, $I$ and $K$ with respect to $\psi_\epsilon$ and $\psi_\eta$ and
finally $\psi_{\epsilon\epsilon}$, $\psi_{\epsilon\eta}$, and
$\psi_{\eta\eta}$ by successive applications of the chain rule:
\begin{eqnarray}
0 & = & {\dLI} \left ( {\partial^2 I \over \partial \epsilon \partial
\psi_\epsilon} + 
{\partial^2 I \over \partial \eta \partial \psi_\eta} \right ) + 
{\dLII}
\left ( \pp{I}{\psi_\eta} \pp{I}{\eta} + \pp{I}{\psi_\epsilon} \pp{I}{\epsilon}
\right ) \\*
 & & ~~~+{\dLK} \left ( {\partial^2 K \over \partial \epsilon \partial
\psi_\epsilon} + 
{\partial^2 K \over \partial \eta \partial \psi_\eta} \right ) + 
{\dLKK}
\left ( \pp{K}{\psi_\eta} \pp{K}{\eta} + \pp{K}{\psi_\epsilon} \pp{K}{\epsilon}
\right ) \nonumber \\*
 & & ~~~+{\dLIK} \left (
\pp{K}{\epsilon} \pp{I}{\psi_\epsilon} +
\pp{I}{\epsilon} \pp{K}{\psi_\epsilon} +
\pp{K}{\eta} \pp{I}{\psi_\eta} +
\pp{I}{\eta} \pp{K}{\psi_\eta}  \right ) \nonumber \\* 
0 & = & a \psi_{\epsilon\epsilon} + b \psi_{\epsilon\eta} + c \psi_{\eta\eta}.
\end{eqnarray}
where
\begin{eqnarray}
a & = & \dLII \left(\pp{I}{\psi_\epsilon}\right)^2+
	\dLKK \left(\pp{K}{\psi_\epsilon}\right)^2+
	\dLI \frac{\partial^2 I}{\partial \psi_\epsilon^2} +
	\dLK \frac{\partial^2 K}{\partial \psi_\epsilon^2} +
	2 \dLIK \pp{I}{\psi_\epsilon} \pp{K}{\psi_\epsilon} \\
b & = & 2 \left [ 
        \dLII \pp{I}{\psi_\eta} \pp{I}{\psi_\epsilon}+
	\dLKK \pp{K}{\psi_\eta} \pp{K}{\psi_\epsilon}+
	\dLI \frac{\partial^2 I}{\partial \psi_\eta \partial\psi_\epsilon}+
	\dLK \frac{\partial^2 K}{\partial \psi_\eta \partial\psi_\epsilon}
	\phantom{\left (\pp{I}{\psi_\epsilon} \pp{K}{\psi_\eta}+
	  \pp{I}{\psi_\eta} \pp{K}{\psi_\epsilon} \right )} \right .
\nonumber \\ *
  &  & ~~~
	\left .
	+ \dLIK \left (\pp{I}{\psi_\epsilon} \pp{K}{\psi_\eta}+
	  \pp{I}{\psi_\eta} \pp{K}{\psi_\epsilon} \right )
	\right ]
\\
c & = & \dLII \left(\pp{I}{\psi_\eta}\right)^2+
	\dLKK \left(\pp{K}{\psi_\eta}\right)^2+
	\dLI \frac{\partial^2 I}{\partial \psi_\eta^2} +
	\dLK \frac{\partial^2 K}{\partial \psi_\eta^2} +
	2 \dLIK \pp{I}{\psi_\eta} \pp{K}{\psi_\eta} 
\end{eqnarray}
Substituting the definitions of $I$ and $K$ in the field geometry yields
\begin{eqnarray}
a & = & \left [ {\bar B}^2 \cos^2 \theta \sin^2 \phi + 4 \psi_\eta
\left ( {\bar B} \cos \theta \sin \phi + \psi_\eta \right ) \right ] \dLII
\nonumber \\*
 & & ~~~ - 4 K {\bar B}^2 \cos^2 \phi \dLKK - 2 {\bar B}^2 \cos^2 \phi \dLK
\nonumber \\*
 & & ~~~ + 16 {\bar B}^2 \cos^2 \phi
\left ( {\bar B} \cos \theta \sin \phi + 2 \psi_\eta \right ) \left ( \psi_\eta
- \psi_\epsilon \right ) \dLIK \\
b & = &  \dLI + 2 \left [ {I \over 2} - \left ( 1 - \cos^2 \theta \sin^2 \phi
\right ) {\bar B}^2 \right ] \dLII \nonumber \\*
 & & ~~~ + 8 K {\bar B}^2 \cos^2 \phi \dLKK + 4 {\bar B}^2 \cos^2 \phi \dLK + 2 K \dLIK \\
c & = & \left [  {\bar B}^2 \cos^2 \theta \sin^2 \phi + 4 \psi_\epsilon
\left ( {\bar B} \cos \theta \sin \phi + \psi_\epsilon \right ) \right ]
\dLII \nonumber \\* 
 & & ~~~ - 4 K {\bar B}^2 \cos^2 \phi \dLKK - 2 {\bar B}^2 \cos^2 \phi \dLK 
\nonumber \\*
 & & ~~~ + 16 {\bar B}^2 \cos^2 \phi
\left ( {\bar B} \cos \theta \sin \phi + 2 \psi_\epsilon \right )
\left ( \psi_\epsilon - \psi_\eta \right ) \dLIK 
\end{eqnarray}
For $\phi=\pi/2$, the equation for the field $\psi$ simplifies considerably,
\begin{eqnarray}
0 & = & \psi_{\epsilon\epsilon} \left [ 4 {\dLII} 
        \left ( \psi_\eta^2 + \psi_\eta {\bar B} \cos \theta + 
        {1 \over 4} {\bar B}^2 \cos^2 \theta \right ) \right ] \\*
\nonumber
 & & ~~ + \psi_{\epsilon\eta} \left [ {\dLI} + 2 {\dLII} 
        \left ( {1 \over 2} I - {\bar B}^2 \sin^2 \theta \right ) \right ] \\*
\nonumber
   & & ~~ + \psi_{\eta\eta} \left [ 4 {\dLII} 
        \left ( \psi_\epsilon^2 + \psi_\epsilon {\bar B} \cos \theta + 
        {1 \over 4} {\bar B}^2 \cos^2 \theta \right ) \right ]
\end{eqnarray}
For $\lag_\rmscr{Classical} = -I/4$, the second derivative of the
Lagrangian with respect to $I$ and the all derivatives with respect to
$K$ are zero.  We obtain $\psi_{\epsilon\eta}=0$ and recover the result
\be
\psi(\epsilon,\eta) = f(\epsilon) + g(\eta) = f(y+t) + g(y-t).
\ee
The terms in this equation correspond physically to ordinary wave
propagation in the negative and positive $y$ directions.

Using the quadratic formula, we can factor the differential equation
\be
\left ( \pp{}{\epsilon} - \rho_+ \pp{}{\eta} \right ) 
\left ( \pp{}{\epsilon} - \rho_- \pp{}{\eta} \right ) \psi = -
\psi_\eta \left ( \pp{}{\epsilon} - \rho_+ \pp{}{\eta} \right ) 
 \rho_-
\label{eq:rhofactor1}
\ee
and
\be
\left ( \pp{}{\epsilon} - \rho_- \pp{}{\eta} \right ) 
\left ( \pp{}{\epsilon} - \rho_+ \pp{}{\eta} \right ) \psi = -
\psi_\eta \left ( \pp{}{\epsilon} - \rho_- \pp{}{\eta} \right ) 
 \rho_+
\label{eq:rhofactor2}
\ee
where
\be
\rho_\pm = { -b \pm \sqrt{b^2-4ac} \over 2 a}.
\ee
Looking at \eqref{rhofactor1} and \eqref{rhofactor2}
we define two new coordinates $(u,v)$ such that
\be
\epsilon = u + v \rmmat{~and~} \eta = -(\rho_+ u + \rho_- v)
\ee
We now define two functions, the Riemann invariants,
\begin{eqnarray}
\psi_+ & = & \psi_\epsilon - \rho_+ \psi_\eta - \int_{v_0}^v \psi_\eta
\left . \pp{\rho_+}{v}\right|_{v=v'} d v'\\*
\psi_- & = & \psi_\epsilon - \rho_- \psi_\eta  - \int_{u_0}^u \psi_\eta
\left . \pp{\rho_-}{u} \right|_{u=u'}d u' 
\end{eqnarray}
which are constant along curves with
\be
\dd{\eta}{\epsilon} = - \rho_\mp(\psi_+,\psi_-;{\bar B}).
\ee
To first order in $\psi_\eta$ and $\psi_\epsilon$, the field variables
of the traveling wave are given in terms of these
functions:
\begin{eqnarray}
E & = & \psi_\eta - \psi_\epsilon = 
  - { (\rho_- - 1) \psi_+ + (\rho_+ - 1) \psi_- \over \rho_- - \rho_+} \\*
B & = & \psi_\eta + \psi_\epsilon =
  { (\rho_- + 1) \psi_+ - (\rho_+ + 1) \psi_- \over \rho_- - \rho_+}
\end{eqnarray}
and the characteristic curves are defined in the $y-t$-plane 
\be
\dd{y}{t} = {1 - \rho_\mp \over 1 + \rho_\mp} = \sigma_\pm.
\ee
Given the specified field configuration, the $\psi_+$ characteristics
travel toward increasing $y$ and the $\psi_-$ ones travel in the
opposite direction.  We can therefore in this
case identify the $\psi_+$ characteristics with the light travel path
\be
\sigma_+ = {1 \over n}
\ee
where $n$ is the index of refraction of the magnetized 
vacuum as discussed in \jcite{Erber}{Erbe66}.

In the limit when the electric and magnetic fields due to the
radiation may be neglected, we get
\begin{eqnarray}
\rho_\pm &=& -x \pm \sqrt{x^2-1} \\*
\sigma_\pm &=& -{x + 1 \pm \sqrt{x^2-1} \over x - 1 \pm \sqrt{x^2-1}}
= \mp \frac{\sqrt{x^2-1}}{x-1} 
\end{eqnarray}
where
\begin{eqnarray}
x &=& \left . \left ( {\dLI} + 2 {\bar B}^2 \cos^2 \theta \sin^2 \phi {\dLII} +
4 {\bar B}^2 \cos^2 \phi {\dLK} \right ) \right / \nonumber \\
& & ~~~ 2 {\bar B}^2 \left ( \cos^2 \theta \sin^2 \phi {\dLII} - 2 \cos^2
\phi {\dLK} \right )
\label{eq:xdef}
\end{eqnarray}
For the weak-field Lagrangian given by \jcite{Berestetskii \etal}{Bere82},
\be
\lag = -{1\over 16 \pi} I + {e^4 \hbar \over 180 \times 8 \pi^2 m^4
c^7} \left ( I^2 - {7 \over 4} K \right).
\ee
This formalism yields an index of refraction,
\be
n = 1 - x^{-1} + {\cal O}(x^{-2}) = 1 + {e^4 \hbar \over 90 \pi m^4
c^7} {\bar B}^2 ( 4 \cos^2 \theta \sin^2 \phi + 7 \cos^2 \phi )
\ee
where we have also used
\be
\dLI \gg {\bar B}^2 \dLII, {\bar B}^2 \dLK.
\ee
This result agrees with \jcite{Erber}{Erbe66} and 
\jcite{Berestetskii \etal}{Bere82}.
}

We characterize the traveling wave by a vector 
potential with one non-zero component,
\be
A_z = \psi(y,t).
\ee
We have assumed that the vector potential has only one independent
component which allows us to treat the problem using characteristics.
Unfortunately, we cannot follow the interaction between two
polarizations which was treated by \jcite{Bialynicka-Birula}{Bial81}.

Using the coordinates, $y$ and $t$, Hamilton's principle assumes
the form
\be
{\partial \over \partial y} \left ( \pp{\lag}{B} \right ) -
{\partial \over \partial t} \left ( \pp{\lag}{E} \right )  = 0.
\ee
where 
\be
\psi_y = B \rmmat{~and~} \psi_t = -E,
\ee
and we have taken $c\equiv 1$.

Since the Lagrangian is a function of $I$ and $K$ alone, we can
rewrite this in terms of derivatives of $\lag$ with respect to $I$ and
$K$, $I$ and $K$ with respect to $\psi_y$ and $\psi_t$ and
finally $\psi_{yy}$, $\psi_{yt}$, and
$\psi_{tt}$ by successive applications of the chain rule.
Taking the partial derivatives yields an equation of the form,
\be
a \psi_{yy} + b \psi_{yt} + c \psi_{tt} = 0
\ee
where
\begin{eqnarray}
a & = & \dLII \left (\pp{I}{B} \right)^2 + \dLKK \left (\pp{K}{B}
\right)^2 + 2 \dLIK \pp{K}{B} \pp{I}{B} + \nonumber \\
 & & ~~~ \dLI \frac{\partial^2 I}{\partial B^2} +
         \dLK \frac{\partial^2 K}{\partial B^2} \\
  & = & 4 \left [ 4  \left ( {\bar B} \cos \theta \sin \phi +
		  B \right)^2 \dLII + \dLI \right ] \\
b & = & -2 \left [ \dLII \pp{I}{E} \pp{I}{B} + \dLKK \pp{K}{E}
\pp{K}{B} + \right . \nonumber \\
 & & ~~~\left . \dLIK \left ( \pp{K}{E} \pp{I}{B} + \pp{K}{B} \pp{I}{E}
\right ) + \dLI \frac{\partial^2 I}{\partial E \partial B} +
\dLK \frac{\partial^2 K}{\partial E \partial B} \right ]\\
  & = & 32 E \left ( {\bar B} \cos \theta \sin \phi + B \right )
\left ( \dLII + 8 \dLIK {\bar B}^2 \cos^2 \phi \right ) \\ 
c & = & \dLII \left (\pp{I}{E} \right)^2 + \dLKK \left (\pp{K}{E}
\right)^2 + 2 \dLIK \pp{K}{E} \pp{I}{E} +   \nonumber \\
 & & ~~~\dLI \frac{\partial^2
I}{\partial E^2} + \dLK \frac{\partial^2 K}{\partial E^2} \\
 & = & 4 \left [ 4 E^2 \dLII - \dLI + \right . \nonumber \\
 &   & ~~~ \left . 8 {\bar B}^2 \cos^2 \phi \left (32
E^2 {\bar B}^2 \cos^2 \phi \dLKK + 8 E^2 \dLIK - \dLK \right ) \right ]
\end{eqnarray}
This equation may be factored yielding
\be
\left ( \pp{}{y} - \tau_+ \pp{}{t} \right )
\left ( \pp{}{y} - \tau_- \pp{}{t} \right ) \psi = - \psi_t \left (
\pp{}{y} - \tau_+ \pp{}{t} \right ) \tau_-
\label{eq:factor1}
\ee
and
\be
\left ( \pp{}{y} - \tau_- \pp{}{t} \right )
\left ( \pp{}{y} - \tau_+ \pp{}{t} \right ) \psi = - \psi_t \left (
\pp{}{y} - \tau_- \pp{}{t} \right ) \tau_+
\label{eq:factor2}
\ee
where
\be
\tau_\pm = -\frac{1}{\sigma_\pm} = \frac{-b \mp \sqrt{b^2-4ac}}{2a}
\ee
and $\sigma_\pm$ is the speed of the travelling wave.

We define two new
coordinates $(u,v)$ such that
\be
y = u + v \rmmat{~and~} t = -(\tau_+ u + \tau_- v)
\ee
We now define the Riemann invariants,
\begin{eqnarray}
\psi_+ & = & \psi_y - \tau_+ \psi_t - \int_{v_0}^v \psi_t
\left . \pp{\tau_+}{v}\right|_{v=v'} d v'\\*
\psi_- & = & \psi_y - \tau_- \psi_t  - \int_{u_0}^u \psi_t
\left . \pp{\tau_-}{u} \right|_{u=u'}d u' 
\end{eqnarray}
which are constant along curves with
\be
\dd{t}{y} = - \tau_\mp(\psi_+,\psi_-;{\bar B}).
\ee

To obtain a cursory understanding of the characteristics of this
equation, we expand each of the coefficients about $E,B = 0$ to first
order yielding,
\begin{eqnarray}
a & = & a_0 + a_B \psi_y - a_E \psi_t + {\cal O}(B^2) \\
b & = & b_0 + b_B \psi_y - b_E \psi_t + {\cal O}(B^2) \\
c & = & c_0 + c_B \psi_y - c_E \psi_t + {\cal O}(B^2).
\end{eqnarray}
The coefficients are
\begin{eqnarray}
a_0 & = & 4 \left [ 4 \left ( {\bar B} \cos \theta \sin \phi \right)^2 \dLII + \dLI
\right ] \\
a_B & = & 16 {\bar B} \cos \theta \sin \phi \left [ 3 \dLII + 4
\left ({\bar B} \cos \theta \sin \phi \right)^2 \dLIII \right ] \\
b_E & = & 32 {\bar B} \cos \theta \sin \phi \left ( \dLII + 8 \dLIK {\bar B}^2
\cos^2 \phi \right ) \\
c_0 & = & -4 \left ( 8 \dLK {\bar B}^2 \cos^2 \phi + \dLI \right ) \\
c_B & = & -\frac{b_E}{2} 
\end{eqnarray}
and
\be
a_E =  b_0 = b_B = c_E = 0.
\ee
Using this linearization we estimate the magnitude of the inhomogeneous
term in \eqref{factor1} and \eqref{factor2} if we assume
$E_t \sim E$ and $B_t \sim B$ 
\be
- \psi_t \left (
\pp{}{y} - \tau_\pm \pp{}{t} \right ) \tau_\mp  = {\cal O} (B^2)
\ee
so we can neglect it to first order.

We estimate the Riemann invariants,
\begin{eqnarray}
\psi_+ & = & \psi_y - \tau_+ \psi_t + {\cal O} (B^2) \\
	& = & B + \tau_+ E + {\cal O} (B^2) \\
\psi_- & = & \psi_y - \tau_- \psi_t + {\cal O} (B^2) \\
	& = & B + \tau_- E + {\cal O} (B^2) 
\label{eq:invEB}
\end{eqnarray}

\begin{figure}
\plotonesc{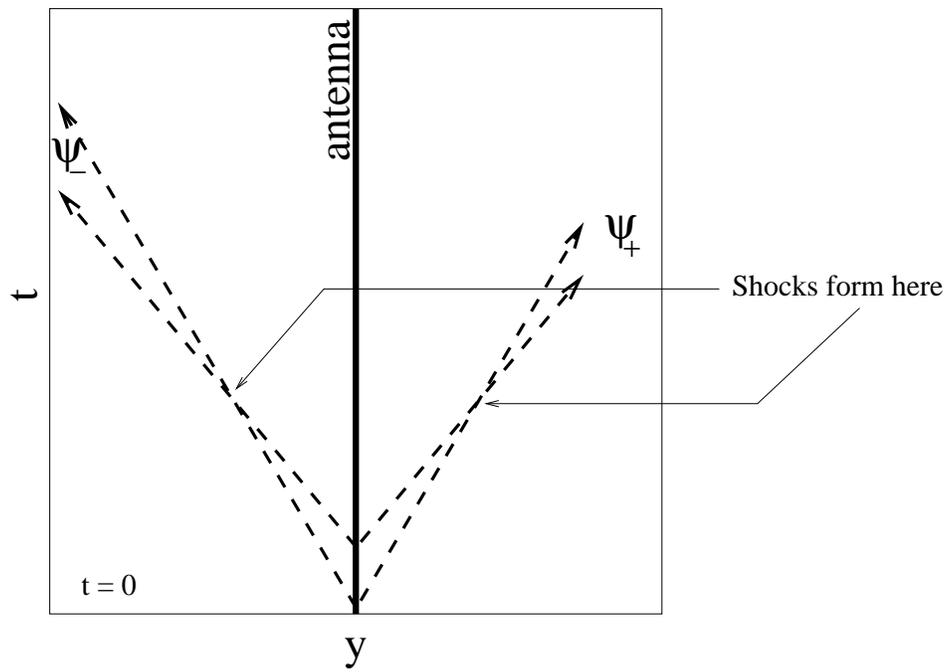}{0.7}
\caption{Characteristics and Shock Formation}
\label{fig:character}
\end{figure}

Now taking the limit where $E$
and $B$ themselves may be neglected we have
\begin{eqnarray}
\sigma_\pm &=& \pm \sqrt{-a_0/c_0} = \pm \left [ \left ( 4 \left ( {\bar B} \cos \theta \sin \phi \right)^2 \dLII + \dLI
\right ) \left / \left ( 8 {\bar B}^2 \cos^2 \phi \dLK  + \dLI \right
. \right ) \right]^{1/2}
\nonumber \\
	& = & \mp \frac{\sqrt{x^2-1}}{x-1}
\label{eq:sigma}
\end{eqnarray}
where
\begin{eqnarray}
x &=& \left . \left ( {\dLI} + 2 {\bar B}^2 \cos^2 \theta \sin^2 \phi {\dLII} +
4 {\bar B}^2 \cos^2 \phi {\dLK} \right ) \right /
2 {\bar B}^2 \left ( \cos^2 \theta \sin^2 \phi {\dLII} - 2 \cos^2
\phi {\dLK} \right )
\label{eq:xdef}
\end{eqnarray}
We can see from this
result that $\tau_+ = - \tau_- + {\cal O}(B)$, therefore we have for
the two Riemann invariants
\be
\psi_\pm = B \pm \tau_+ E + {\cal O}(B^2)
\ee
\figref{character} depicts how two adjacent characteristic may
intersect.  Additionally, we see that $\psi_-$ characteristics (\ie
lines along which $\psi_-$ is constant) that originate from regions
without wave fields cross the $\psi_+$ characteristics.  Therefore, we
can argue that $\psi_-=B-\tau_+ E + {\cal O}(B^2) = 0$
throughout the region to the
right of the antenna because the fields are zero in this region.  We can use the same argument for the region to
the left of the antenna and find that in general
$B=\tau_\pm E + {\cal O}(B^2)$ along
the $\psi_\pm$ characteristics.
Furthermore, to first order,
both $\psi_\pm$ are constant along the characteristics; therefore, the
slopes of the characteristics which depend only on $\psi_+,\psi_-$ and
the constant background field must be constants and the
characteristics travel at a constant speed.

Using the figure as a guide, we estimate the distance over which
two adjacent
characteristics can travel before intersecting is given by
\be
\Delta y  =  c \left ( \pp{\tau_\pm}{t} \right )^{-1} =
\frac{-2 a_0 \tau_\pm c}{\left(a_B \tau_\pm^2 + c_B\right)B_
t - 2 c_B \tau_\pm E_t} .
\ee
Here we have used that fact that $\psi_+$ characteristics
travel with velocity $-\tau_-=\tau_+ + {\cal O}(B)$.

To work with this equation further, we define an opacity due to shock
formation and use $B = \tau_\pm E$.
We obtain
\begin{eqnarray}
\kappa &=& (\Delta y)^{-1} = \frac{a_B c_0 + c_B a_0}{2 a_0^2 \tau_\pm
c} B_t \\
&=& \pm \frac{a_B c_0 + c_B a_0}{2 a_0^2} \sqrt{-\frac{a_0}{c_0}}
\frac{B_t}{c} \\
&=& \pm
8 {\bar B}_B \left ( 4 \dLII {\bar B}_B^2 + \dLI \right)^{-3/2} \Biggr [
\dLII \dLI
+ \left ( 2 \dLI \dLIK + 6 \dLII \dLK \right)  {\bar B}_E^2
\nonumber \\*
& &~~~
+ \left (\dLI \dLIII  + \left(\dLII\right)^2 \right) {\bar B}_B^2
+ 8 \left ( \dLIII \dLK +\dLII  \dLIK \right ) {\bar B}_E^2 {\bar B}_B^2
\Biggr ]
\nonumber \\*
& &~~~
\times \left(8 \dLK {\bar B}_E^2 + \dLI \right)^{-1/2} \frac{B_t}{c}
\label{eq:kappapart}
\end{eqnarray}
where we have defined
\be
{\bar B}_B = {\bar B} \cos\theta \sin\phi \rmmat{~and~} {\bar B}_E =
{\bar B} \cos\phi.
\ee
Where two of these characteristic curves intersect, a shock will form.
To move further, we must choose the proper Lagrangian.

\section{The Non-Linear Lagrangian}

\jcite{Heisenberg and Euler}{Heis36} and \jcite{Weisskopf}{Weis36}
independently derived the effective Lagrangian of the
electromagnetic field using electron-hole theory.
\jcite{Schwinger}{Schw51} later rederived the same result using
quantum electrodynamics.  In rationalized electromagnetic units, the
Lagrangian is given by
\begin{eqnarray}
\lag & = & -{1 \over 4} I + \lag_1 \\
\lag_1 &=& {e^2 \over h c} \int_0^\infty e^{-\zeta} 
{\d \zeta \over \zeta^3} \left \{ i \zeta^2 {\sqrt{-K} \over 4} \times
{ \cos \left ( {\zeta \over E_k} \sqrt{-{I\over 2} + i {\sqrt{-K}\over 2}} \right ) +
\cos \left ( {\zeta \over E_k} \sqrt{-{I\over 2} - i {\sqrt{-K}\over 2}} \right ) \over
\cos \left ( {\zeta \over E_k} \sqrt{-{I\over 2} + i {\sqrt{-K}\over 2}} \right ) -
\cos \left ( {\zeta \over E_k} \sqrt{-{I\over 2} - i {\sqrt{-K}\over 2}} \right ) } 
 + |E_k|^2 + {\zeta^2 \over 6} I \right \}.
\end{eqnarray}
where $E_k=B_k={m^2 c^3 \over e \hbar}$.  In the weak field limit
Heisenberg and Euler give
\be
\lag \approx -{1 \over 4} I + E_k^2 {e^2 \over h c} \left [
{1 \over E_k^4} \left ( {1 \over 180} I^2 - {7 \over 720} K \right
) + {1 \over E_k^6} \left ( {13 \over 5040} K I - {1 \over 630}
I^3 \right ) \cdots \right ]
\label{eq:heweak}
\ee
We define a dimensionless parameter $\xi$ to characterize the field
strength ($I$)
\be
\xi = {1 \over E_k} \sqrt{I \over 2}
\ee
and use the analytic expansion of this Lagrangian for small $K$
derived by \jcite{Heyl and Hernquist}{Heyl96a}:
\be
\lag_1 = \lag_1(I,0) + K \left . \pp{\lag_1}{K} \right |_{K=0} +
\frac{K^2}{2} \left . \frac{\partial^2 \lag_1}{\partial K^2} \right
|_{K=0} + \cdots
\label{eq:lag1exp}
\ee
The first two terms of this expansion are given by
\begin{eqnarray}
\lag_1(I,0) & = & {e^2 \over h c} \frac{I}{2}
X_0\left(\frac{1}{\xi}\right) \\
 \left . \pp{\lag_1}{K} \right |_{K=0} & = & {e^2 \over h c}
\frac{1}{16 I} X_1\left(\frac{1}{\xi}\right) 
\end{eqnarray}
where
\begin{eqnarray}
X_0(x) & = & 4 \int_0^{x/2-1} \ln(\Gamma(v+1)) \d v
+ \frac{1}{3} \ln \left ( \frac{1}{x} \right )
+ 2 \ln 4\pi - (4 \ln A+\frac{5}{3} \ln 2) \nonumber \\
& & ~~ - \left [ \ln 4\pi + 1 +  \ln \left ( \frac{1}{x} \right ) \right ] x
+ \left [ \frac{3}{4} + \frac{1}{2} \ln \left ( \frac{2}{x} \right )
\right ]
x^2
\label{eq:x0anal} \\
X_1(x) & = & - 2 X_0(x) + x X_0^{(1)}(x) + \frac{2}{3} X_0^{(2)} (x) -
\frac{2}{9} \frac{1}{x^2}
\label{eq:x1anal} 
\end{eqnarray}
and
\begin{eqnarray}
X_0^{(n)}(x) &=& \frac{\d^n X_0(x)}{\d x^n} \\
\ln A &=& \frac{1}{12} - \zeta^{(1)}(-1) \approx 0.2488.
\end{eqnarray}
where $\zeta^{(1)}(x)$ is the first derivative of the Riemann
Zeta function.

Using these definitions we can derive the various partial derivatives
important for shock formation
\begin{eqnarray}
\dLI &=& -\frac{1}{4} + \frac{e^2}{h c} \Biggr [
\frac{1}{2} X_0\left(\frac{1}{\xi}\right)
- \frac{1}{4} X_0^{(1)}\left(\frac{1}{\xi}\right) \xi^{-1} \Biggr ] \\
\dLK &=& \frac{1}{288} \frac{e^2}{h c} E_k^{-2} \Biggr [
-2 - \left ( 18  X_0\left(\frac{1}{\xi}\right)
-6 X_0^{(2)}\left(\frac{1}{\xi}\right) \right ) \xi^{-2}
+ 9  X_0^{(1)}\left(\frac{1}{\xi}\right) \xi^{-3} \Biggr ] \\
\dLII &=&  \frac{1}{16} \frac{e^2}{h c} E_k^{-2} \Biggr [
- X_0^{(1)}\left(\frac{1}{\xi}\right) \xi^{-3}
+ X_0^{(2)}\left(\frac{1}{\xi}\right) \xi^{-4}
\Biggr ] \\
\dLIK &=& \frac{1}{384} \frac{e^2}{h c} E_k^{-4} \Biggr [
\left ( 12  X_0\left(\frac{1}{\xi}\right)
-4 X_0^{(2)}\left(\frac{1}{\xi}\right) \right ) \xi^{-4}
-\left ( 3 X_0^{(1)}\left(\frac{1}{\xi}\right)
+2 X_0^{(3)}\left(\frac{1}{\xi}\right) \right ) \xi^{-5}
-3 X_0^{(2)}\left(\frac{1}{\xi}\right)\xi^{-6} \Biggr ] \\
\dLIII &=& \frac{1}{64} \frac{e^2}{h c} E_k^{-4} \Biggr [
3 X_0^{(1)}\left(\frac{1}{\xi}\right) \xi^{-5}
- 3  X_0^{(2)}\left(\frac{1}{\xi}\right)  \xi^{-6}
- X_0^{(3)}\left(\frac{1}{\xi}\right)  \xi^{-7} \Biggr ]
\end{eqnarray}

\section{The Opacity to Shocking}

Using the results of the previous section we can expand the opacity
($\kappa$) to order $e^2/h c$, which results in a substantial
simplification
\ba
\kappa &=& -32 \frac{B_t {\bar B}_B}{c} \left ( \dLII
+ \dLIII {\bar B}_B^2 + 2 \dLIK {\bar B}_E^2\right )
+ {\cal O} \left [ \left(\frac{e^2}{hc}\right)^2\right]
\label{eq:kappalag} \\ 
       &=& - \frac{B_t {\bar B}_B}{c B_k^2} \frac{e^2}{h c} 
\Biggr \{ 
2 \left [
- X_0^{(1)}\left(\frac{1}{\xi}\right) \xi^{-3}
+ X_0^{(2)}\left(\frac{1}{\xi}\right) \xi^{-4}
\right ]
+ \frac{1}{2} \left(\frac{{\bar B}_B}{B_k}\right)^2 \Biggr [
3 X_0^{(1)}\left(\frac{1}{\xi}\right) \xi^{-5}
- 3  X_0^{(2)}\left(\frac{1}{\xi}\right)  \xi^{-6}
- X_0^{(3)}\left(\frac{1}{\xi}\right)  \xi^{-7} \Biggr ]
\nonumber \\*
& & ~~~
+ \frac{1}{6} \left(\frac{{\bar B}_E}{B_k}\right )^3 \Biggr [
\left ( 12  X_0\left(\frac{1}{\xi}\right)
-4 X_0^{(2)}\left(\frac{1}{\xi}\right) \right ) \xi^{-4}
-\left ( 3 X_0^{(1)}\left(\frac{1}{\xi}\right)
+2 X_0^{(3)}\left(\frac{1}{\xi}\right) \right ) \xi^{-5}
-3 X_0^{(2)}\left(\frac{1}{\xi}\right)\xi^{-6}
\Biggr ]
\Biggr \} 
\nonumber \\*
& & ~~~
+ {\cal O} \left [ \left(\frac{e^2}{hc}\right)^2\right]
\ea
where we have focussed on the propagation of the $\psi_+$
characteristics.  The results for the $\psi_-$ characteristics are
identical in magnitude and follow from symmetry.

From the form of \eqref{kappapart}, we see that the opacity is zero
for waves traveling with their magnetic field vectors perpendicular
to the external field ($\perp$ mode).  This result agrees with
Bialynicka-Birula's analysis \cite{Bial81} who found that although a
wave in the $\perp$ mode readily generates waves in the $\|$ mode, a
wave in the $\perp$ mode does not change to first order.  These
selection rules result from the $CP$ invariance of QED and may be
gleamed from the selection rules for photon splitting \cite{Adle71}.

Because our analysis tracks the evolution of only a single
mode, we will calculate the opacity in the limit where the magnetic 
field of the wave is parallel to the external field.  Waves in the
$\|$ mode generate higher harmonics in the $\|$ mode but none in the
$\perp$ mode.  In this limit,
\be
{\bar B}_B = {\bar B} \rmmat{~and~} {\bar B}_E = 0
\ee
and \eqref{kappapart} simplifies further,
\be
\kappa =  \frac{1}{2} \frac{e^2}{h c}
\left [ X_0^{(1)}\left(\frac{1}{\xi}\right) \xi^{-2}
-X_0^{(2)}\left(\frac{1}{\xi}\right) \xi^{-3}
+X_0^{(3)}\left(\frac{1}{\xi}\right) \xi^{-4} \right ]
\frac{1}{c}\frac{B_t}{B_k}  
+ {\cal O} \left [ \left(\frac{e^2}{hc}\right)^2\right]
\ee
We define a dimensionless auxiliary function $F(\xi)$ to characterize
the opacity due to shocking
\be
\kappa = - F(\xi) \frac{1}{c} \frac{B_t}{B_k} = - F(\xi) l_B^{-1}.
\label{eq:kappaf}
\ee
We define $l_B$ to be the characteristic length over which the
magnetic field of the wave would change by $B_k$.  $l_B$ is positive
for sections of the wave where the magnetic field strength increases
as its passes a stationary observer, or equivalently in the frame of
the wave itself, $l_B$ is negative in sections where the field
strength decreases in the direction of propagation.

The function $F(\xi)$ may be expanded in the weak-field
limit ($\xi<0.5$) yielding
\ba
F(\xi) &=& -4 \frac{e^2}{hc} \frac{1}{\xi} \sum_{j=1}^\infty 2^{2j} B_{2(j+1)}
\frac{j+1}{2j+1} \xi^{2j}
+ {\cal O} \left [ \left(\frac{e^2}{hc}\right)^2\right] \\
\label{eq:fexpanw}
       &=& \frac{e^2}{h c} \left ( \frac{16}{45} \xi - \frac{32}{35}
\xi^2 + \cdots \right )
+ {\cal O} \left [ \left(\frac{e^2}{hc}\right)^2\right] 
\ea
where $B_n$ denotes the $n$th Bernoulli number.  In the strong-field
limit ($\xi>0.5$), we obtain
\be
F(\xi) = \frac{e^2}{hc}
\left \{ \frac{2}{3} \frac{1}{\xi} + \frac{1}{2} \left ( \ln\pi - 2 -
\ln\xi \right ) \frac{1}{\xi^2} +
\frac{2}{\xi} \sum_{j=3}^\infty \frac{(-1)^{j-1}}{2^j}
\frac{4(j-1)-j^2}{j-1} \zeta(j-1) \xi^{-j} \right \}
+ {\cal O} \left [ \left(\frac{e^2}{hc}\right)^2\right]
\ee
where $\zeta(x)$ is the Riemann Zeta function and we have used
the expansions of \jcite{Heyl and Hernquist}{Heyl96a}.

\begin{figure}
\plotonesc{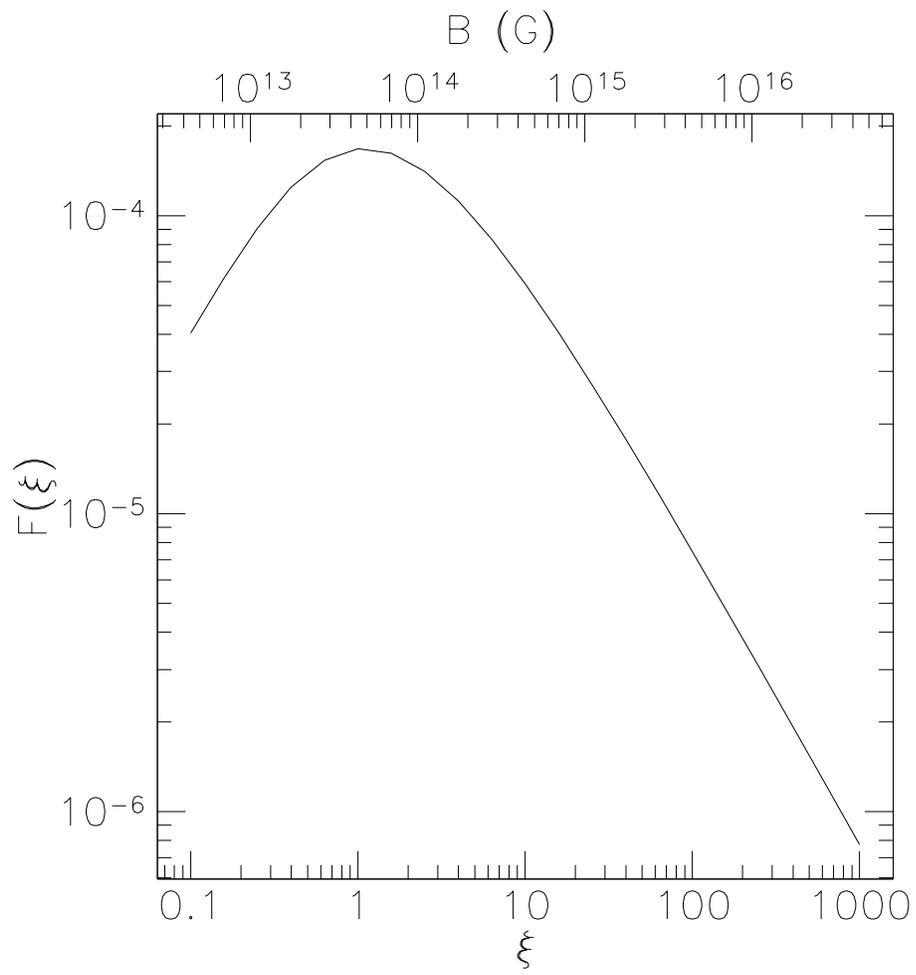}{0.7}
\caption{The figure depicts the auxiliary function $F(\xi)$ as a
function of $\xi$}
\label{fig:fshock}
\end{figure}

Note from \figref{fshock} that $F(\xi)$ is positive for all field
strengths.
Thus, from examination of \eqref{kappaf}, we see that the opacity is
positive (shocks will develop) in regions where the magnetic field is
increasing toward the direction of propagation in the frame of the
wave.  $F(\xi)$ also reaches a maximum near the critical field
strength.

\section{The Physical Shock: Jump Conditions}

\begin{figure}
\plotonesc{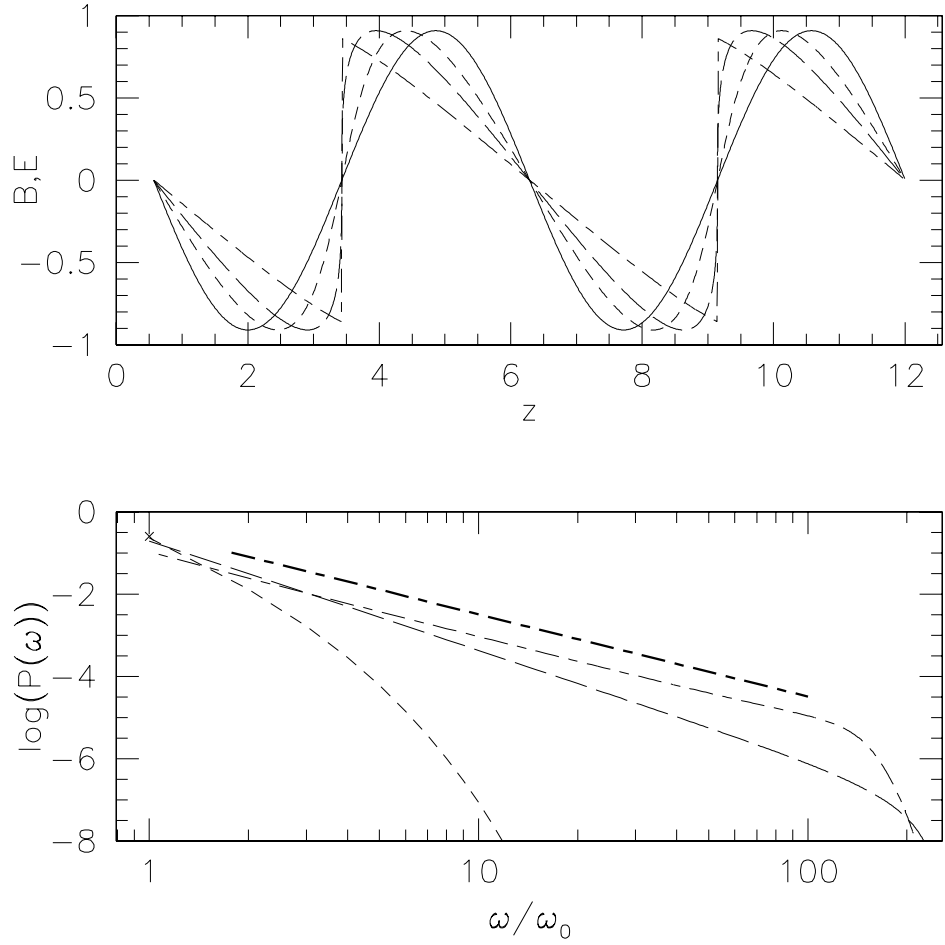}{0.7}
\caption{Evolution of a wave toward shock formation in the frame of the wave.
The upper panel traces the wave in the comoving frame.  The lower panel traces
the power spectrum of the wave.
The successive lines denote original wave, the wave at an optical depth of
one-half, at an optical depth of one and at an optical depth of two.  
The bold line shows a power spectrum of $\nu^{-2}$.}
\label{fig:waveshock}
\end{figure}

We expect a shock to develop when and where the value of the Riemann
invariant is discontinuous.  From \eqref{invEB} and using the result
$B=\tau_\pm E$, we see that the invariants are simply the electric and
magnetic field strengths associated with the wave.  \figref{waveshock}
schematically depicts the evolution of the wave. The shock begins to
form at an optical depth of one where the field of the wave becomes
discontinuous. 

As in fluid shocks, dissipative processes prevent the field strengths
from becoming double valued.  We use the Maxwell equal area prescription
(\cite{Land6}) to calculate the shock profile after the characteristic
analysis indicates that the field strengths become double valued.  We start 
with a sinusodial wavefront,
\be
B(y,t) = - B_0 \sin( y - \sigma_+ t) = -B \sin \upsilon_0
\ee
and obtain the following equation for the characteristics
\be
\upsilon(\tau) = \upsilon_0 + \tau \sin \upsilon_0
\label{eq:wavechar}
\ee
in the frame of the wave.  Furthermore, for convenience we use the optical depth 
to shock formation as the time unit.

From \figref{waveshock} and \eqref{wavechar} we see that the wave evolves 
symmetrically about $\upsilon=\pi$.  The position of the shock is given by the 
location which divides the double-valued regions into equal areas.  By symmetry 
this occurs at $\upsilon=\pi$.  The wavefront at $\tau=2$ is constructed in this 
manner.

To determine the dissipation of energy by the shock, we calculate the mean
power of the wave 
\ba
P &=& \frac{1}{\pi} \int_0^\pi c B^2 \d \upsilon =
    \frac{1}{\pi} \sigma_+ B_0^2 \int_0^\pi \sin^2 \upsilon_0 \d \upsilon \\
  &=& \frac{B_0^2}{\pi} \int_{\upsilon=0}^{\upsilon=\pi} 
    \sin^2 \upsilon_0 \left ( 1 + \tau \cos \upsilon_0 \right ) \d \upsilon_0 \\
  &=& \frac{B_0^2}{\pi} \left ( 
    \frac{\upsilon_{0,s}}{2} - \frac{1}{4} \sin 2\upsilon_{0,s} 
    + \frac{1}{3} \tau \sin^3 \upsilon_{0,s} \right )
\ea
where $\upsilon_{0,s}$ is the smallest solution of
\be
\pi = \upsilon_{0,s} + \tau \sin \upsilon_{0,s}
\label{eq:upshock}
\ee
That is, the shock is located at $\upsilon_s=\pi$. 
For $\tau\leq 1$ the only real solution to \eqref{upshock} is 
$\upsilon_{0,s} = \pi$.  Therefore, before the shock forms the mean power in the
wave is simply $1/2 \sigma_+ B_0^2$.

Unfortunately, in general this equation can only be solved numerically; however,
two limits exist which can be treated analytically.  As the shock just begins to 
develop $\upsilon_0 \approx \upsilon$ near the shock, so
$\upsilon_{0,s} \approx \pi$.  If we expand \eqref{upshock} about 
$\upsilon_{0,s} = \pi$, we obtain
\ba
\upsilon_{0,s} &\approx& \sqrt{\frac{6 {\tau - 1}}{\tau}} \\
P &=& \sigma_+ B_0^2\left [ \frac{1}{2} - 
      \frac{8}{5 \pi} \sqrt{6} (\tau-1)^{5/2} 
    + \frac{45}{7 \pi} \sqrt{6} (\tau-1)^{7/2} + {\cal O} (\tau-1)^{9/2} \right ]
\ea
We find that as soon as the shock forms at $\tau=1$, the shock begins to dissipate
energy from the wave.  Additionally, the dissipation does not begin abruptly.  
This first two terms in this expansion are accurate to 
$\sim 1\%$ for $\tau - 1 \lesssim 0.2$.

\begin{figure}
\plotonesc{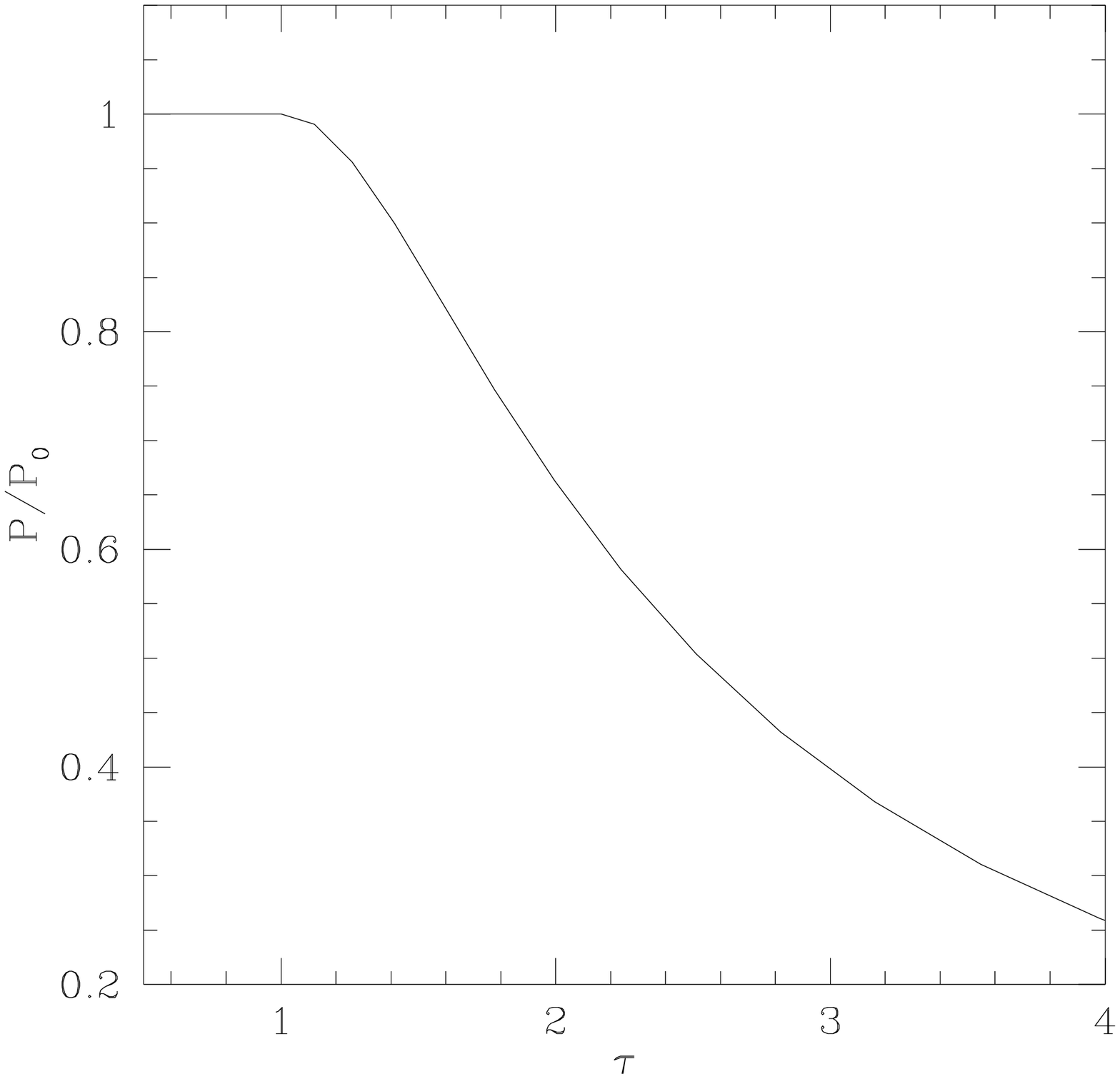}{0.5}

\plotonesc{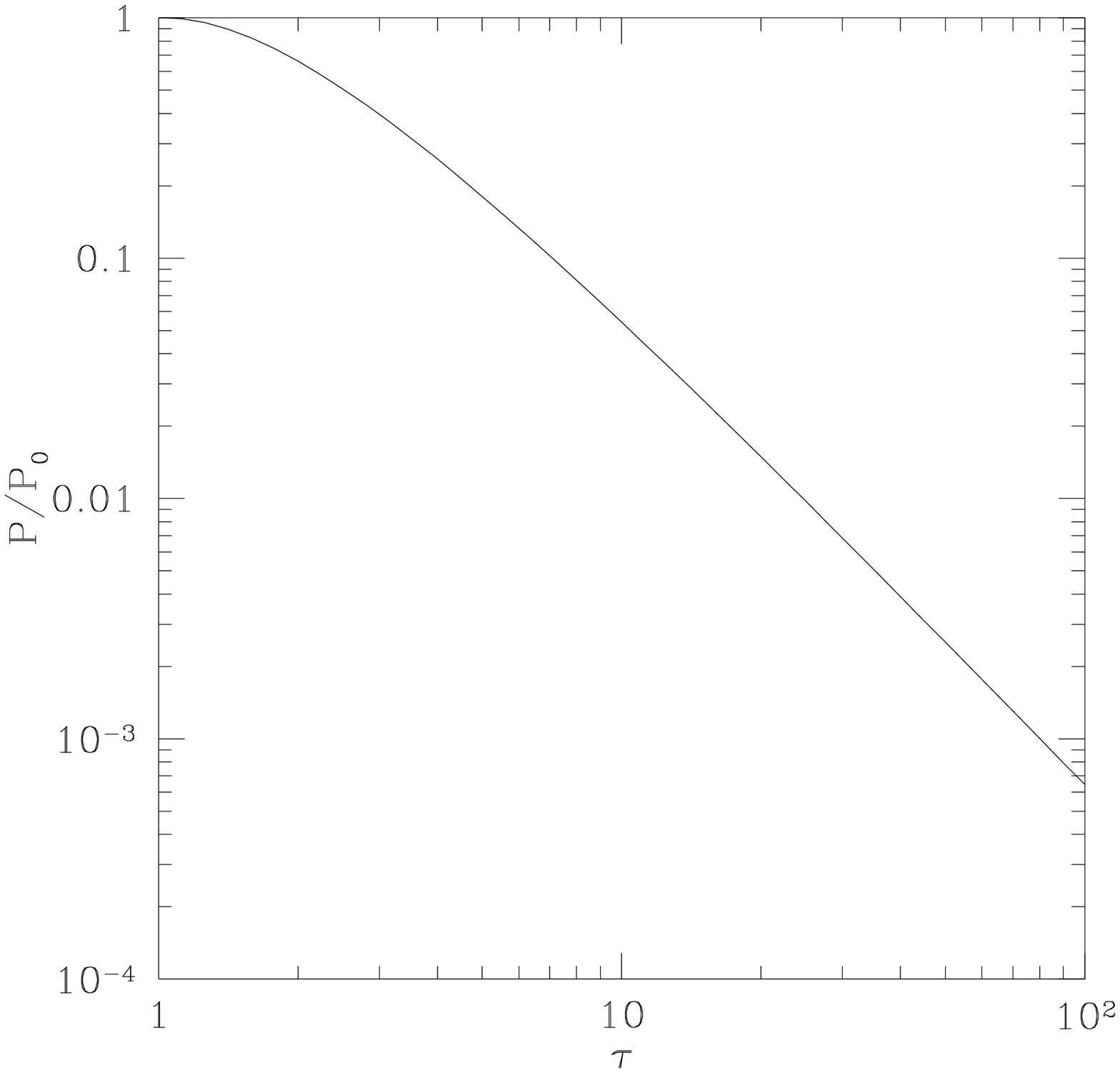}{0.5}
\caption{The evolution of the power carried by the wave before and after
shock formation.}
\label{fig:powev}
\end{figure}

At late times, we can find a solution to \eqref{upshock} such that 
$\upsilon_s \approx 0$.  Here we obtain
\ba
\upsilon_{0,s} &\approx& \frac{\pi}{\tau+1} \\
P &\approx& \frac{\pi^2}{3} \frac{1}{(\tau+1)^2}.
\ea
This expression is accurate to one percent for $\tau > 12$.  The upper panel
of \figref{powev} depicts the energy dissipation soon after the shock forms.  
It is apparent that the dissipation begins smoothly.  The lower panel shows the
late evolution.

Here, in the preceding analysis we have
assumed that the field gradients are small both in our linearization
and in our selection of the Heisenberg-Euler Lagrangian.  Our
linearization, specifically the assumption that the gradient of the
fields is small relative to the fields breaks down when
\be
\lambda_e \pp{B}{z} \sim B
\ee
where $\lambda_e$ is the Compton wavelength of the electron.  The
Heisenberg-Euler Lagrangian also breaks down when the field changes
by $B_k$ over scales similar to $\lambda_e$.  In this limit, one must
use more powerful techniques, such as Schwinger's proper-time method
\cite{Schw51} to determine the effective Lagrangian.

When the field changes dramatically over scales similar to
$\lambda_e$, our fluid approximation breaks down. We estimate 
the thickness of the shock to be approximately $\lambda_e$.
To further understand the properties of the shock, we derive the jump 
conditions across the shock discontinuity.  

We move to the rest-frame of the shock and insist on the continuity of
the dual to the field tensor and the energy-momentum tensor:
\be
\partial_\nu {\cal F}^{\mu\nu} = 0 \rmmat{~and~}
\partial_\nu \Theta^{\mu\nu} = 0
\ee
where
${\cal F}^{\mu\nu}=\frac{1}{2} \epsilon^{\mu\nu\delta\gamma}
F_{\delta\gamma}$ (\eg \cite{Land2}) and
\be
\partial_\nu \equiv \pp{}{x^\nu}.
\ee
The first condition follows from the gauge invariance
of the fields. The second condition represents the conservation of
energy and momentum.  These jump conditions are equivalent to those
used by \jcite{Boillat}{Boil72} who insisted that the
dual of field tensor be continuous and that the Euler-Lagrange condition
be satisfied.

For clarity, in contrast to the analysis of the
preceding sections, we examine the energy-momentum and the field
tensors of the combined wave and constant background field.
Using the techniques outlined in \cite{Itzy80}, we find
the energy momentum tensor for non-linear electrodynamics.  The
canonical tensor (${\tilde{\Theta}}^{\mu\nu}$)
is constructed from the Lagrangian by means of a Legendre transformation,
\be
{\tilde{\Theta}}^{\mu\nu} = \pp{\lag}{\left [\partial_\mu A_\rho\right]}
\partial^\nu A_\rho - g^{\mu\nu} \lag
\ee
where $A_\rho$ is the potential four-vector of the electromagnetic
field.  We construct the more familiar symmetrized
energy-momentum tensor ($\Theta^{\mu\nu}$) by subtracting a total divergence,
\ba
\Theta^{\mu\nu} &=& \pp{\lag}{\left [\partial_\mu A_\rho\right]}
\partial^\nu A_\rho - g^{\mu\nu} \lag - \partial_\rho
\left ( \pp{\lag}{\left [\partial_\mu A_\rho\right]} A^\nu \right ) \\
&=& \pp{\lag}{\left [\partial_\mu A_\rho\right]}
\partial^\nu A_\rho - g^{\mu\nu} \lag
-  \pp{\lag}{\left [\partial_\mu A_\rho\right]} \partial_\rho A^\nu
- A^\nu \partial_\rho
\pp{\lag}{\left [\partial_\mu A_\rho\right]} 
\label{eq:theta3} \\
&=& \pp{\lag}{\left [\partial_\mu A_\rho\right]}
F^\nu_{\phantom{\nu}\rho}
- g^{\mu\nu} \lag
\ea
where the final term of \eqref{theta3} is zero by the Euler-Lagrange
condition.  Using the definitions of $I$ and $K$ we obtain,
\be
\Theta^{\mu\nu} = \left ( 4 \dLI F^{\mu\rho} - 8 J \dLK {\cal
F}^{\mu\rho} \right ) F^\nu_{\phantom{\nu}\rho} - g^{\mu\nu} \lag.
\ee
where $J = F_{\mu\nu} {\cal F}^{\mu\nu}$.

Simplifying this expression yields,
\be
\Theta^{\mu\nu} = -4 \dLI F^{\mu\rho}
F_\rho^{\phantom{\rho}\nu} - g^{\mu\nu} \left ( \lag - 2 K \dLK \right). 
\ee
which for $\lag = -\frac{1}{4} F_{\rho\sigma} F^{\rho\sigma}$, the linear
case, yields
\be
\Theta^{\mu\nu} = \frac{1}{4} g^{\mu\nu} F_{\rho\sigma} F^{\rho\sigma}
+ F^{\mu\rho} F_\rho^{\phantom{\rho}\nu}.
\ee
in agreement with Itzykson and Zuber's result \cite{Itzy80}.

When determining shock jump conditions for a fluid, one calculates the
velocity of the discontinuity relative to the rest frame of the fluid.
In analogy to a fluid,  we can associate the rest frame of the
electromagnetic field  with the frame in which the energy-momentum
tensor is diagonal.  This frame exists if either of the two invariants
($I$ and $K$) is non-zero.  In this frame, the electric and magnetic
fields are parallel and their magnitudes are given by
\ba
B^2 &=& \frac{1}{4} \left (  I \pm \sqrt{I^2-K} \right ), \\
E^2 &=& \frac{1}{4} \left ( -I \pm \sqrt{I^2-K} \right ) 
\ea
where the $+$ sign is chosen for $I>0$ and $-$ for $I<0$.  If we take
the fields to point along the $x$-axis, we obtain
\ba
\Theta^{00} = -\Theta^{11} &=& -4 \dLI E^2 - \lag + 2 K \dLK, \\
\Theta^{22} = \Theta^{33} &=& -4 \dLI B^2 + \lag - 2 K \dLK.
\ea
To apply this to the geometry of the previous sections
(\figref{geometry}), we will assume that $K=0$ and that 
the shock front is parallel to the $x-z$ plane and traveling
toward increasing $y$.

The geometry leads to the jump conditions:
\be
[\Theta^{02}] = 0, [\Theta^{22}]=0 \rmmat{~and~}
[{\cal F}^{12}]=0.
\ee
across the shock.
We calculate the components of the energy-momentum tensor in the
shock frame through a boost.  If $K\neq 0$, the Lorentz transformation
from the rest frames to the shock frame would include a
rotation as well.
By boosting the rest frames into the shock frame we get the
following jump conditions
\begin{eqnarray}
[\Theta^{02}] &=& \left [2 \gamma^2 v \dLI I  \right] = 0 \\
\protect{[\Theta^{22}]} &=& \left [ \lag-2 \gamma^2 \dLI I \right] = 0 \\
\protect{[{\cal F}^{12}]} &=& [ \sqrt{I/2} \gamma v ] = 0  
\end{eqnarray}
where $v$ is the speed that the diagonalizing frame is moving relative
to the shock and $\gamma=\left (1-v^2 \right)^{-1/2}$.
The three conditions are
physically conservation of energy and momentum flux, and the continuity
of the electric field parallel to the
surface of shock.  Analogous jump conditions are given by \cite{Land6}
for relativistic fluid shocks.  If we define,
\be
e = w-p = -\lag,~p=\lag - 2 \dLI I,~w = -2 \dLI I \rmmat{~and~} n=-\sqrt{I}
\ee
we obtain the following jump conditions
\begin{eqnarray}
[\Theta^{02}] &=& \left [\gamma^2 v w \right] = 0 \\
\protect{[\Theta^{22}]} &=& \left [p + \gamma^2 v^2 w \right] = 0 \\
\protect{[{\cal F}^{12}]} &=& [\gamma v n] = 0 
\end{eqnarray}
Landau and Lifshitz \cite{Land6} give the velocities of the
diagonalizing frames relative to the discontinuity 
\begin{eqnarray}
v_1 &=&
\sqrt{\frac{p_2-p_1}{e_2-e_1}\frac{e_2+p_1}{e_1+p_2}}
\label{eq:v1} \\
v_2 &=&
\sqrt{\frac{p_2-p_1}{e_2-e_1}\frac{e_1+p_2}{e_2+p_1}}
\label{eq:v2}
\end{eqnarray}
and the equation of the shock adiabatic
\be
\left ( \frac{w_1}{n_1} \right )^2 - \left ( \frac{w_2}{n_2} \right )^2
+ (p_2 - p_1) \left ( \frac{w_1}{n_1^2} + \frac{w_2}{n_2^2} \right ) = 0
\ee
where the subscripts 1 and 2 denote conditions on either side of the shock.
The equation for the shock adiabatic is automatically satisfied to first 
order.

Taking \eqref{v1} and \eqref{v2} and assuming that the shock strength is a
linear perturbation on the background field we get
\ba
v_{1,2} &=& \sqrt{\left(2 I \dLII + \dLI\right)\Biggr/\dLI} \left \{ 1
- \frac{1}{2} \Delta I \Biggr / \dLI 
\left [  (1\pm 1) \dLII + I \dLIII\right ] \right \} + {\cal O} \left (
\frac{B^2}{\bar B^2} \right), \\
        &=& \sigma_+ \left [ 1 + \left ( 1 \pm 1 \right ) 
\frac{1 - \sigma^2}{4} \frac{\Delta I}{I} 
- \frac{1}{2} I \Delta I  \dLIII \Biggr / \dLI 
\right ]
+ {\cal O} \left (\frac{B^2}{\bar B^2} \right), \\
        &=& \sigma_+ \left [ 1 + \left ( 1 \pm 1 \right ) 
\frac{1 - \sigma^2}{2} \frac{\Delta B}{\bar B} 
- 4 {\bar B}^3 \Delta B  \dLIII \Biggr / \dLI 
\right ]
+ {\cal O} \left (\frac{B^2}{\bar B^2} \right), 
\ea
where $\Delta I=I_1-I_2>0, \Delta B=B_1-B_2>0$ and
we have used \eqref{sigma} to simplify the expression.

It is more useful to find the speed at which the shocks travel
relative to the external field.  In general, the
rest frame travels at a velocity that satisfies \cite{Land2}
\be
\frac{{\bf V}}{1+|\bf V|^2} =
\frac{\efi \times \left (\bfi + {\bar \bfi} \right )}
{|\efi|^2+|\bfi + {\bar \bfi}|^2}
\ee
For $K=0$, the rest frame of the electromagnetic field travels at
\be
v_{1,2} = \mp \frac{E}{\bar B} +
{\cal O} \left ( \frac{B^2}{\bar B^2} \right ) = \mp \sigma \frac{B}{\bar B} +
{\cal O} \left ( \frac{B^2}{\bar B^2} \right )
\ee
in the $y$-direction relative to external field behind and ahead of the
shock, respectively.

By summing the velocities of the rest frames relative to the external
field and the shock relative to the rest frames, we find that the 
shock travels at the speed of light in the medium relative to the
external field
\be
v_{1,2} = \sigma
+ {\cal O} \left ( \frac{B^2}{\bar B^2} \right)
\ee
so we find in agreement with \jcite{Boillat}{Boil72} that the
Heisenberg-Euler Lagrangian does not admit shocks to first order in the
strength of the discontinuity, if shocks are strictly defined as
discontinuities that travel at a speed other that the speed of
``sound'' in the medium.  

\section{Conclusions}

We have developed a relativistic fluid dynamic description of the
electromagnetic field to derive the characteristic equations for
electromagnetic waves in the presence of a strong external magnetic
field.  By using a analytic expressions for the effective Lagrangian of
QED, we have obtained simple expressions to estimate the opacity of
waves to shocking for arbitrary magnetic field strengths.  Furthermore,
by deriving a concise expression for the energy-momentum tensor for an
arbitrary electromagnetic Lagrangian, we have calculated the shock jump
conditions and find that the discontinuities travel at the speed of
light in the medium relative to the external magnetic field.  By
extending this fluid dynamic description, we follow the eventual decay
of a electromagnetic disturbance travelling through an intense magnetic
field. 

For shocks to form from electromagnetic waves, not only is a strong
external field required but also a source of coherent radiation.  A
prime location for electromagnetic shocks is the vicinity of a neutron
star with field strengths approaching and exceeding the critical value
and coherent electromagnetic Alfven waves.  The study of the nonlinear
corrections to the propagation of radiation through a plasma is beyond
the scope of this work; however, we expect shock formation to be a
hallmark of the nonlinear corrections of quantum electrodynamics and
possibly an important process in the energy transmission near neutron
stars.

\acknowledgements

This material is based upon work supported under a National Science
Foundation Graduate Fellowship. L.H. thanks the National Science
Foundation for support under the Presidential Faculty Fellows
Program.  J.S.H. also acknowledges Cal Space grant CS-12-97 and a 
Lee A. DuBridge Postdoctoral Scholarship.

\begin{thebibliography}{10}

\bibitem{Heis36}
W. Heisenberg and H. Euler, Z. Physik {\bf 98},  714  (1936).

\bibitem{Weis36}
V.~S. Weisskopf, Kongelige Danske Videnskaberns Selskab, Mathematisk-Fysiske
  Meddelelser {\bf 14},  1  (1936).

\bibitem{Lutz59}
M. Lutzky and J.~S. Toll, Phys. Rev. {\bf 113},  1649  (1959).

\bibitem{Zhel82}
V.~V. Zheleznyakov and A.~L. Fabrikant, Sov. Phys. JETP {\bf 55},  794  (1982).

\bibitem{Bial81}
Z. Bialynicka-Birula, Physica D {\bf 2},  513  (1981).

\bibitem{Schw51}
J. Schwinger, Physical Review {\bf 82},  664  (1951).

\bibitem{Heyl96a}
J.~S. Heyl and L. Hernquist, Phys. Rev D, {\bf 55}, 2449 (1997).

\bibitem{Adle71}
S.~L. Adler, Ann. Phys. {\bf 67},  599  (1971).

\bibitem{Land2}
L.~D. Landau and E.~M. Lifshitz, {\em The Classical Theory of Fields}, fourth
  ed. (Pergamon, Oxford, 1987).

\bibitem{Boil72}
G. Boillat, Phys. Lett. {\bf 40A},  9  (1972).

\bibitem{Itzy80}
C. Itzykson and J.-B. Zuber, {\em Quantum Field Theory} (McGraw-Hill, New York,
  1980).

\bibitem{Land6}
L.~D. Landau and E.~M. Lifshitz, {\em Fluid Mechanics}, 2nd  ed. (Pergamon,
  Oxford, 1987).

\end{thebibliography}
\end{document}